\documentclass[english,journal=jctcce,manuscript=article,articletitle=true,layout=twocolumn]{achemso}
\usepackage[T1]{fontenc}
\usepackage[latin9]{inputenc}
\usepackage{color}
\usepackage{array}
\usepackage{url}
\usepackage{multirow}
\usepackage{amsmath}
\usepackage{amssymb}
\usepackage{graphicx}
\usepackage[numbers]{natbib}

\makeatletter

\title{Linear weak scalability
of density functional theory calculations without imposing electron
localization}
\author{Marcel D. Fabian}
\affiliation{Fritz Haber Research Center for Molecular Dynamics and the Institute
of Chemistry, The Hebrew University of Jerusalem, Israel}
\author{Ben Shpiro}
\affiliation{Fritz Haber Research Center for Molecular Dynamics and the Institute
of Chemistry, The Hebrew University of Jerusalem, Israel}
\author{Roi Baer}
\email{roi.baer@huji.ac.il}
\affiliation{Fritz Haber Research Center for Molecular Dynamics and the Institute
of Chemistry, The Hebrew University of Jerusalem, Israel}
\providecommand{\tabularnewline}{\\}

\numberwithin{equation}{section}

\usepackage{chemformula}
\SectionNumbersOn

\makeatother

\usepackage{babel}
\begin{document}
\begin{abstract}
Linear scaling density functional theory (DFT) approaches to the electronic structure of materials are often based on the tendency of electrons to localize in large atomic and molecular systems. However, in many cases of actual interest, such as semiconductor nanocrystals, system sizes can reach a substantial extension before significant electron localization sets in, causing a considerable deviation from linear scaling. Here, we address this class of systems by developing a massively parallel DFT approach which doesn't rely on electron localization and is formally quadratic scaling yet enables highly efficient linear wall-time complexity in the weak scalability regime. The method extends from the stochastic DFT approach described in Fabian et al. WIRES:Comp. Mol. Science, e1412 2019 but is entirely deterministic. It uses standard quantum chemical atom-centered Gaussian basis sets to represent the electronic wave functions combined with Cartesian real-space
grids for some operators and enables a fast solver for the Poisson equation. Our
main conclusion is that when a processor-abundant high-performance
computing (HPC) infrastructure is available, this type of approach
has the potential to allow the study of large systems in regimes where
quantum confinement or electron delocalization prevents linear-scaling.
\end{abstract}

\section{\label{sec:Introduction}Introduction}

In the past few decades, the supercomputers' massive number-crunching
power, measured in floating-point operations per second (FLOPS), has
grown a million-fold \citep{heldens2020thelandscape} and is currently
pushing towards the exaflop ($10^{18}$ FLOPS) realm. Combining this
new technology with electronic structure calculations can revolutionize
computational materials science and biochemistry, provided we complement
it with algorithms that can efficiently exploit its massively parallel-based
infrastructure. 

One of the key questions then becomes how to quantify the efficiency
of a certain algorithm on a massively parallel machine. A crucial
measure in this regard is the \textit{speedu}\emph{p, }which we define
as the ratio
\begin{equation}
\mathbb{S}\left(W,M\right)\equiv\frac{T_{1}\left(W\right)}{T_{M}\left(W\right)}\label{eq:Define-speedup}
\end{equation}
between the wall-times, $T_{1}\left(W\right)$ for executing a given
computational work $W$ using a single processor and $T_{M}\left(W\right)$
for its execution using $M$ processors working in parallel. In operational
regimes where the speedup is nearly proportional to $M$, i.e. $\mathbb{S}=\mathbb{E}\times M$
there is a clear advantage in using a parallel multiprocessor approach
where $\mathbb{E}$ is the efficiency, with $\mathbb{E}=1$ being
ideal. 

The efficient use of parallel computing was discussed by Amdahl in
his seminal paper \citep{amdahl1967validity}, where he identified
in $W$ an inherently serial (subscript $s$) and parallelizable (subscript
$p$) part, $W=W_{s}+W_{p}$. He assumed that the execution wall-time
is independent of $M$ for completing $W_{s}$ and decreases linearly
with $M$ for $W_{p}$. Amdahl defined the \emph{serial fraction} as
$s_{A}=\frac{T_{1}\left(W_{s}\right)}{T_{1}\left(W\right)}$, measured
on a single processor machine for a given job independent of $M$.
With this definition, the speedup can be expressed as:  $\mathbb{S_{A}}\left(W,M\right)=\left(s_{A}+\frac{1-s_{A}}{M}\right)^{-1}$
(\emph{Amdahl's law},\emph{ }also called\emph{ strong scalability})
and saturates once $M$ exceeds the value of $1/s_{A}$.

Gustafson pointed out \citep{gustafson1988reevaluating,gustafson1988development}
that in real-world usage the definition for the serial
fraction should dependent on $M$, due to the fact, that one does
not generally take a fixed-sized problem, as Amdahl did, but rather
scales the workload $W$ with the available computing power. He then
defined the serial fraction $s_{G}=\frac{T_{M}\left(W_{s}\right)}{T_{M}\left(W\right)}$
as measured on the $M$-processor system and showed that the speedup
can be expressed as $\mathbb{S_{G}}\left(M\right)=s_{G}+M\left(1-s_{G}\right)$
(\emph{Gustafson's law} also called \emph{weak scalability}), enabling
linear speed up which does not inherently saturate as $M$ increases. 

These considerations can be applied to electronic structure calculations
of extended systems in DFT codes that lower the cubic scaling by taking
advantage of electron localization \citep{mohr2014daubechies,nakata2020largescale,kuhne2020cp2kan,olsen2020daltonproject,garcia2020siestarecent,prentice2020theonetep,rudberg2018ergoan,goedecker1999linearscaling,yang1995adensitymatrix,galli2000largescale,scuseria1999linearscaling,baer1997sparsity,baer1998energyrenormalizationgroup,baer2013selfaveraging,osei-kuffuor2014accurate2,cytter2018stochastic,fabian2019stochastic,li2019stochastic,chen2019energywindow,chen2019overlapped,chen2021stochastic}.
 For linear-scaling schemes, the Amdahl serial fraction $s_{A}=\frac{T_{1}\left(W_{s}\right)}{T_{1}\left(W\right)}$
is expected to be system-size independent (since both timings in the
numerator and the denominator scale linearly with system size) while
for codes of higher algorithmic complexity, $s_{A}$ decreases as
system size increases \citep{corsetti2014performance}. In a weak
scalability analysis of the linear scaling codes Gustafson's serial
fraction $s_{G}=\frac{T_{M}\left(W_{s}\right)}{T_{M}\left(W\right)}$
is also expected to be system-size independent (since both timings
in the numerator and the denominator scale linearly with system size)
and therefore take to form: $s_{G}=\left(1+\frac{M_{0}}{M}\right)^{-1}$,
where $M_{0}$ is a constant (depending on the hardware, algorithm).
For large $M$, the speedup saturates to $\mathbb{S_{G}}\left(M\right)\to1+M_{0}$,
but if $M_{0}$ is very large there is a sizable regime where $M\ll M_{0}$
and the $s_{G}$ is essentially zero\emph{ }so \emph{an ideal linear
speedup} emerges, as reported, for example, for the CONQUEST code
\citep{arita2014largescale,nakata2020largescale}, even up to $M=200,000$
cores on the Fujitsu-made K-computer. It is clear from the previous
studies mentioned above that it is important to determine the strong
and weak scalability properties of codes that can use massively parallel
machines, because they are sensitive to many details concerning hardware,
systems size, algorithmic scaling etc. 

In this paper we develop an efficiently parallelizable, (semi)local
DFT approach which offers quadratic scaling with system size and does
not involve approximations derived from assuming electron localization.
It combines several approaches, such as atom-centered
Gaussian basis sets and real-space grids for providing the electrostatic
and exchange-correlation energies (similar to SIESTA \citep{garcia2020siestarecent}
and CP2K/Quickstep \citep{kuhne2020cp2kan}) as well as Chebyshev
expansion techniques for representing the density matrix \citep{goedecker1994efficient,goedecker1995tightbinding,goedecker1995lowcomplexity,baer1997sparsity}.
We describe the theory and implementation in section~\ref{sec:Method},
where we also provide an illustration of the non-localized nature
of electrons in the large benchmarking systems we use (see Figure~\ref{fig:Sparsity}).
Next, we present the algorithmic complexity and the parallel strong/weak
scalability properties of our approach in section~\ref{sec:Scaling-properties-of},
and finally, we summarize and discuss the conclusions in section \ref{sec:Summary-and-Conclusions}.

\section{\label{sec:Method}Method}

In our method, we work with standard quantum chemistry basis sets,
composed of atom-centered local functions $\phi_{\alpha}\left(\boldsymbol{r}\right)$,
$\alpha=1,\dots K$. For calculating the necessary integrals, solving
the Poisson equations, and generating the exchange-correlation potentials,
we use a 3D Cartesian real-space grid of equidistant points spanning
a simulation box, containing the system's atoms and electronic density.
For this purpose, we developed an efficient method for evaluating
the basis functions on a relevant set of grid points, outlined in
section A of the supplementary material. Our method of combining basis
functions and real-space grids is similar in spirit to those existing
in literature, such as SIESTA \citep{garcia2020siestarecent} and
CP2K/Quickstep \citep{kuhne2020cp2kan}, but differs in important
details. Unlike SIESTA, we use standard non-orthogonal Gaussian basis
sets and unlike Quickstep we represent the basis functions on the
grid where all integrals are performed as summations. The first type
of integral that we have to evaluate on the grid then, is the overlap
matrix:
\begin{equation}
S_{\alpha\beta}=h^{3}\sum_{\boldsymbol{g}}\phi_{\alpha}\left(\boldsymbol{r_{g}}\right)\phi_{\beta}\left(\boldsymbol{r_{g}}\right),\label{eq:Smat}
\end{equation}
where $\boldsymbol{r_{g}}$ are the grid points and $h$ is the grid-spacing.
Next, the kinetic energy integrals are evaluated as
\begin{equation}
T_{\alpha\beta}=\frac{1}{2}h^{3}\sum_{\boldsymbol{g}}\nabla\phi_{\alpha}\left(\boldsymbol{r_{g}}\right)\cdot\nabla\phi_{\beta}\left(\boldsymbol{r_{g}}\right),\label{eq:Tmat}
\end{equation}
where the derivatives of the basis functions are calculated analytically
and then placed on the grid (see section A.2.3 of the supplementary
material for details). To avoid an excessive number of grid points,
the equally-spaced grid is complemented with norm-conserving pseudopotentials
\citep{troullier1991efficient}, representing the effects of the tightly
bound core electrons (which are not treated explicitly) and taken
into account in the KS Hamiltonian, represented by the Fock matrix
\begin{equation}
F^{KS}=T+V^{NL}+V^{KS},\label{eq:FockMatrix}
\end{equation}
where 
\begin{equation}
V_{\alpha\beta}^{NL}=h^{3}\sum_{\boldsymbol{g}}\phi_{\alpha}\left(\boldsymbol{r_{g}}\right)\sum_{C\in nuclei}\hat{v}_{nl}^{C}\:\phi_{\beta}\left(\boldsymbol{r_{g}}\right)
\end{equation}
are the integrals for the non-local pseudopotential and 
\begin{equation}
V_{\alpha\beta}^{KS}=h^{3}\sum_{\boldsymbol{g}}\phi_{\alpha}\left(\boldsymbol{r_{g}}\right)v_{KS}\left(\boldsymbol{r_{g}}\right)\phi_{\beta}\left(\boldsymbol{r_{g}}\right)\label{eq:VKSmat}
\end{equation}
are the KS potential integrals, where: 
\begin{align}
v_{KS}\left(\boldsymbol{r_{g}}\right) & =\sum_{C\in nuclei}v_{loc}^{C}\left(\boldsymbol{r_{g}}-\boldsymbol{R}_{C}\right)\label{eq:VKS(rg)}\\
 & \,\,\,\,\,\,\,\,\,\,\,+v_{H}\left[n\right]\left(\boldsymbol{r_{g}}\right)+v_{xc}\left[n\right]\left(\boldsymbol{r_{g}}\right).\nonumber 
\end{align}
In Eq.~(\ref{eq:VKS(rg)}), $v_{H}\left[n\right]\left(\boldsymbol{r_{g}}\right)$
is the Hartree potential on the grid which is evaluated directly from
the grid representation of the electron density $n\left(\boldsymbol{r_{g}}\right)$
by a reciprocal space-based method for treating long range interactions
\citep{martyna1999areciprocal}. The exchange-correlation potential
$v_{xc}\left[n\right]\left(\boldsymbol{r}_{g}\right)$ (within the
local density approximation (LDA)) is also determined on the grid
directly from the electron density. From the grid representation of
the pseudopotentials\footnote{Here we use the Kleinman-Bylander (KB) form \citep{kleinman1982efficacious},
which produces two types of operators, a non-local potential operator
$\hat{v}_{nl}^{C}\equiv v_{nl}^{C}\left(\boldsymbol{r}-\boldsymbol{R}_{C},\boldsymbol{r}'-\boldsymbol{R}_{C}\right)$
which is defined in a small sphere around each atomic core ($\boldsymbol{R}_{C}$
is the location of atom $C$) and a scalar potential $v_{loc}^{C}\left(\boldsymbol{r}-\boldsymbol{R}_{C}\right)$
containing the long-range electron-shielded nucleus Coulomb attraction.} we obtain the potential $v_{loc}^{C}\left(\boldsymbol{r_{g}}-\boldsymbol{R}_{C}\right)$
appearing in Eq.~(\ref{eq:VKS(rg)}) for nucleus $C$ at position
$\boldsymbol{R}_{C}$ and, by grid integration, the matrix $V^{NL}$
appearing in Eq.~(\ref{eq:FockMatrix}). All integral calculations
are performed in parallel for different basis function pairs; for
more details see the supplementary material C.

The electron density on the grid is formally defined as
\begin{equation}
n\left(\boldsymbol{r_{g}}\right)=2\sum_{\alpha,\beta}^{K}P_{\alpha\beta}\:\phi_{\alpha}\left(\boldsymbol{r_{g}}\right)\phi_{\beta}\left(\boldsymbol{r_{g}}\right),\label{eq:density}
\end{equation}
where $P$ is the density matrix (DM) and the factor of two comes
from integration over spin degrees of freedom. The DM must obey an
electron conserving criterion, namely that the integral over all grid
points evaluates to the total number of electrons in the system: $h^{3}\sum_{\boldsymbol{g}}\:n\left(\boldsymbol{r}_{\boldsymbol{g}}\right)=N_{e}$.
Indeed, performing this integral and using Eq.~(\ref{eq:Smat}) and
(\ref{eq:density}) we find 
\begin{equation}
N_{e}=2\text{Tr}\left[PS\right].\label{eq:Ne}
\end{equation}
This relation is part of a more general requirement, that the Kohn-Sham
eigenstates are populated according to the Fermi-Dirac function $f_{FD}\left(\varepsilon\right)=\frac{1}{1+e^{\beta\left(\varepsilon-\mu\right)}}$
where $\varepsilon$ is the corresponding energy eigenvalue. For the
DM, this condition can be satisfied by defining \citep{fabian2019stochastic}:
\begin{align}
P & =f_{FD}\left(S^{-1}F^{KS}\right)S^{-1}.\label{eq:RDM}
\end{align}
For finite-temperature DFT, $\beta$ is the inverse temperature and
$\mu$ is the chemical potential. For ground-state calculations $\beta$
obeys $\beta\left(\varepsilon_{L}-\varepsilon_{H}\right)\gg1$, where
$\varepsilon_{L}$ ($\varepsilon_{H}$) is the Kohn-Sham eigenvalue
of the lowest unoccupied (highest occupied) molecular orbital. The
chemical potential in the Fermi-Dirac function is adjusted to reproduce
the systems' number of electrons $N_{e}$ through Eq.~(\ref{eq:Ne}).

\begin{figure*}
\includegraphics[scale=0.25]{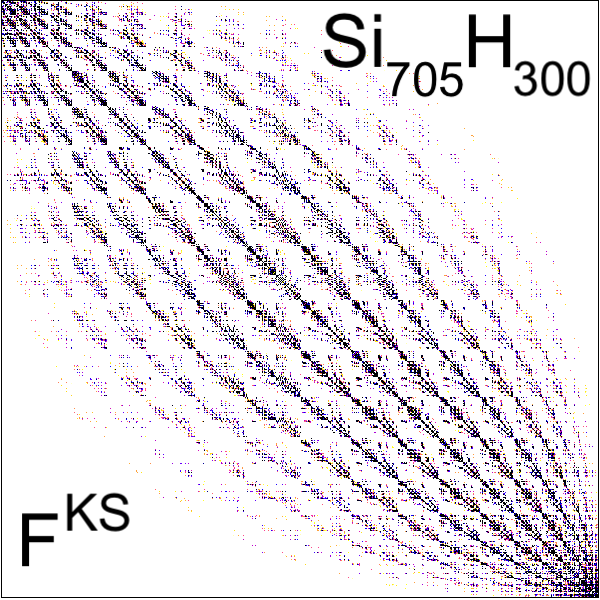}\includegraphics[scale=0.25]{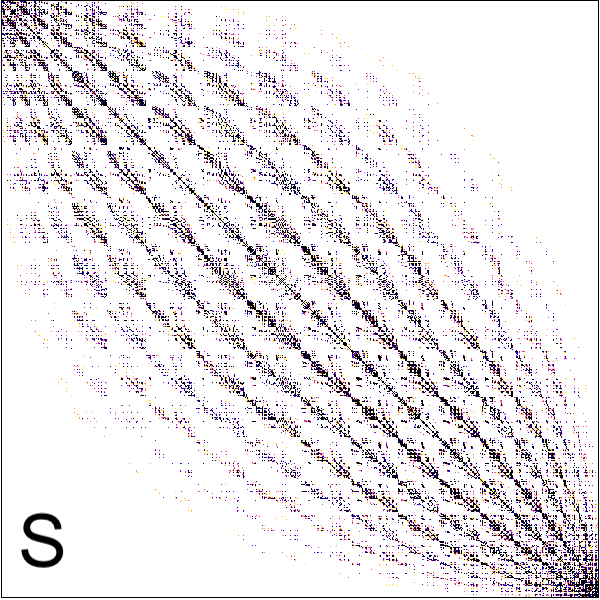}\includegraphics[scale=0.25]{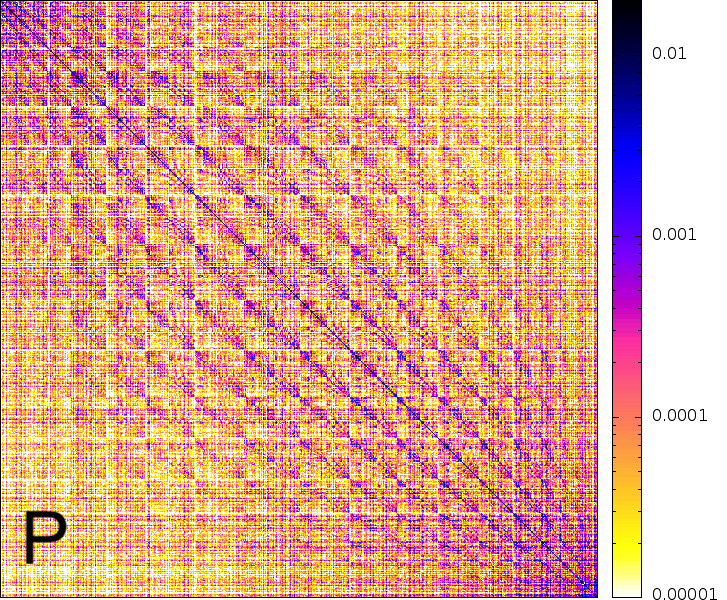}

\includegraphics[scale=0.25]{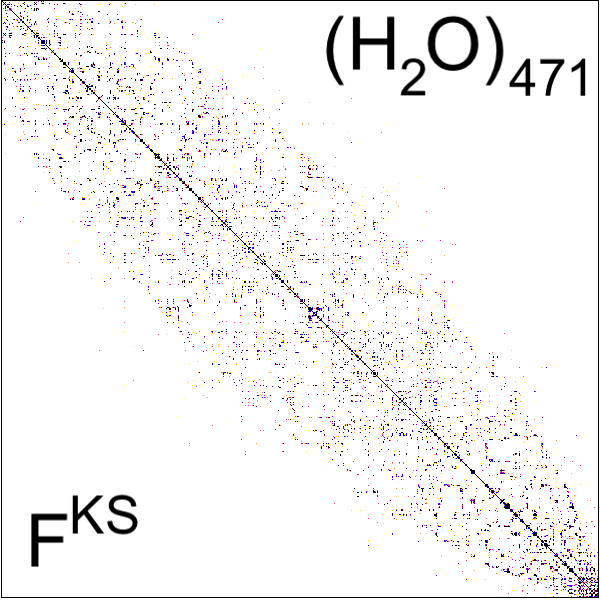}\includegraphics[scale=0.25]{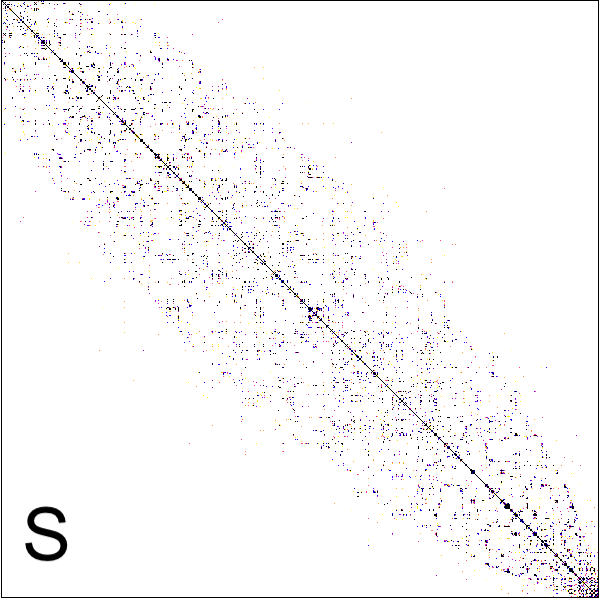}\includegraphics[scale=0.25]{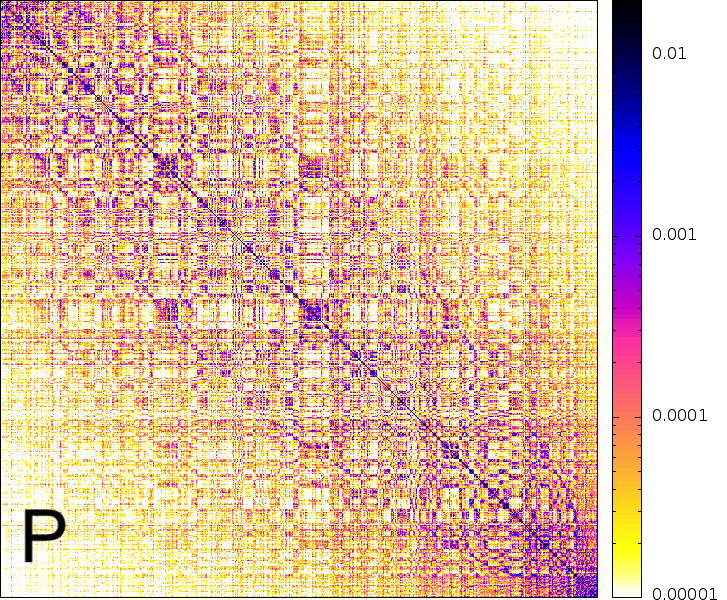}

\caption{\label{fig:Sparsity}The Fock ($F^{KS}$), overlap ($S$) and DM ($P$)
matrices of \ch{Si705H300} and \ch{(H2O)471} cluster (calculated
within the LDA) are shown using a color-coded plot. The basis set
in both systems is similar in size, with $K\approx6000$. For clearer
inspection of the sparsity pattern the rows and columns are permuted
so as to achieve a minimum bandwidth around the diagonal. We applied
the \protect\url{MinimumBandWidthOrdering} command of Mathematica\textsuperscript{\tiny\textregistered}
\citep{incmathematica} to the atomic proximity matrix $D_{AB}=\Theta\left(R_{0}-R_{AB}\right)$
(where $R_{AB}$ is the distance between any pair of atoms $A$, $B$
and $R_{0}=10a_{0}$ is the proximity distance), giving a permutation
which is then used to order the atom-centered basis functions. }
\end{figure*}
\begin{figure}
\centering{}\includegraphics[width=1\columnwidth]{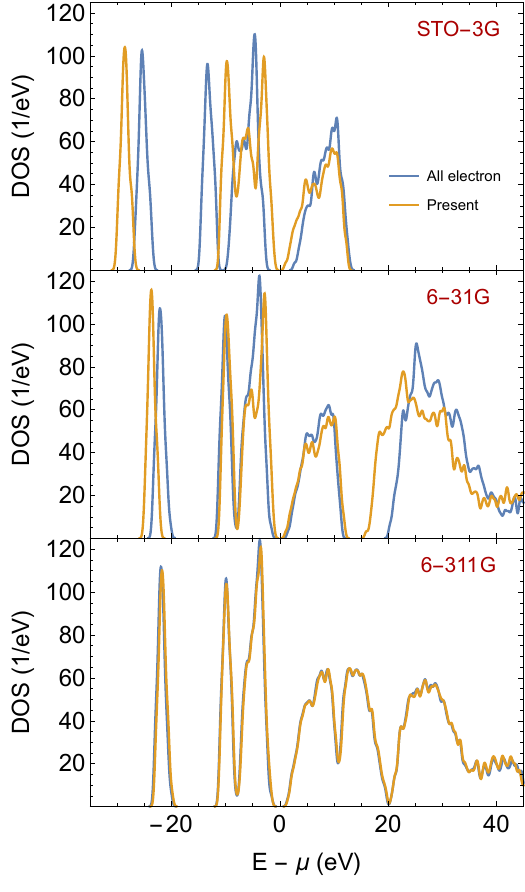}\caption{\label{fig:ComparisonDOS}The density of state (DOS)
as a function of energy, shifted by the chemical potential $\mu$,
for a water cluster (H$_{2}$O)$_{100}$ in the LDA at a fixed geometry.
The three panels compare all-electron calculations (performed by Q-Chem
\citep{shao2015advances}) with valence-electron-only calculations
(performed by the present approach, using pseudopotentials). Each
panel presents the results for different Gaussian basis sets, from
single to triple zeta quality (STO-3G, 6-31G and 6-311G). Both calculations
use the eigenvalues $\varepsilon_{n}$ of the converged KS Hamiltonian
to obtain the DOS function $\rho_{DOS}\left(\varepsilon\right)=2\times\frac{1}{\sqrt{2\pi}\sigma}\sum_{n}e^{-\frac{\left(\varepsilon-\varepsilon_{n}\right)^{2}}{2\sigma^{2}}}$
where $\sigma=0.01E_{h}$. The calculation in the present approach
used $\beta=100E_{h}^{-1}$ with a real-space grid of spacing $\Delta x=0.33\,a_{0}$. }
\end{figure}

The use of atom-centered local basis functions allows for sparsity
in the basic matrices $F^{KS}$ and $S$, as illustrated in Figure~\ref{fig:Sparsity}
for two systems of similar size but different chemical nature, a $2.5\text{nm}$
(diameter) semiconductor nanocrystal \ch{Si705H300} and a $3\text{nm}$
water cluster \ch{(H2O)471}. For the matrix representation in Figure~\ref{fig:Sparsity},
we have ordered the atoms (and the basis functions associated with
them) in a way that takes into account their spatial proximity (near
atoms tend to have similar indices). Therefore, it is clear by mere
inspection that $F^{KS}$ and $S$ have a relatively small spatial
range and are therefore quiet sparse. Our approach makes an effort
to exploit this property by using sparse matrix algebra. Despite
the spatial locality of $F^{KS}$ and $S$, $P$ in these large systems
is highly non-local, expressing the physical fact, that the electronic
coherence in these systems is long ranged. For the silicon system,
this fits our intuition, namely that silicon is by nature a semiconductor,
with properties which are close to those of metals. Although water
is a large band-gap system, it is known that under LDA it exhibits
very small HOMO-LUMO gaps \citep{herbert2005calculation,rudberg2012difficulties,lever2013electrostatic,sosavazquez2015sizedependent}
(see also Figure~\ref{fig:ComparisonDOS}).

The various expectation values of relevant observables (i.e., operators
in the grid representation) can be expressed as trace operations:
\begin{equation}
\left\langle \hat{O}\right\rangle =2\textrm{Tr}\left[PO\right]\label{eq:<O>}
\end{equation}
where
\begin{equation}
O_{\alpha\beta}=h^{3}\sum_{\boldsymbol{g}}\phi_{\alpha}\left(\boldsymbol{r_{g}}\right)\hat{O}\phi_{\beta}\left(\boldsymbol{r_{g}}\right)
\end{equation}
is the matrix representation of the one body operator $\hat{O}$ in
the atomic basis. In order to expedite the calculation we need to
parallelize the computational work, and this can be done by representing
the trace operations as a sum over unit column vectors $u_{\alpha}$
(with coordinates $\left(u_{\alpha}\right)_{\beta}=\delta_{\alpha\beta}$,
i.e., zeros in all positions except at $\alpha$), computed column
by column: 
\begin{equation}
\left\langle \hat{O}\right\rangle =2\sum_{\alpha}^{K}u_{\alpha}^{T}OPu_{\alpha}.\label{eq:col-by-col}
\end{equation}
For achieving this, we treat the DM as an operator, i.e. we devise
a linear-scaling method for applying it to the column vector $u_{\alpha}$,
based on Eq.~(\ref{eq:RDM}): $Pu_{\alpha}=f_{FD}\left(S^{-1}F^{KS}\right)S^{-1}u_{\alpha}$.
The operation $S^{-1}u_{\alpha}$ is performed by the linear-scaling
preconditioned conjugate-gradient approach involving repeated application
of the sparse overlap matrix $S$ on column vectors\footnote{\label{fn:PCG}We use the incomplete Cholesky preconditioning \citep{scott2014hslmi28}
for the conjugated gradient approach implemented in the HSL-MI28 and
MI21 codes, respectively, where HSL is a collection of FORTRAN codes
for large scale scientific computation ( http://www.hsl.rl.ac.uk/ accessed on March 5 2022).}. The operation of $f_{FD}\left(S^{-1}F^{KS}\right)$ on the column
vector $S^{-1}u_{\alpha}$ employs a Chebyshev expansion \citep{goedecker1994efficient,baer1997chebyshev}
of the function $f_{FD}\left(\varepsilon\right)$, which results in
repeated applications of the operator $S^{-1}F^{KS}$ to column vectors.
Details are described in the supplementary material B. The entire
procedure can be readily distributed over several processors in parallel,
each commissioned with a distinct set of $u_{\alpha}$ column vectors.
This calculation method has the additional benefit that it avoids
storage of the non-sparse DM. We discuss the algorithmic complexity
of the approach, as well as its weak and strong scalability in section~\ref{sec:Scaling-properties-of}. 

Equations (\ref{eq:Smat})-(\ref{eq:RDM}) and the techniques of their
application discussed above form a series of nonlinear equations that
must be solved together, to give the self-consistent-field (SCF) solution.
The procedure is iterative and uses the direct inversion of the iterative
subspace (DIIS) convergence acceleration method \citep{pulay1982improved}.
Once converged various expectation values such as charges and multipoles,
density of states and polarizability can be calculated, as well as
forces on the nuclei \citep{shpiro2021forcesfrom}, which can be used
for structure optimization.

In order to check and validate the implementation
of the algorithm outlined above, we show in Figure~\ref{fig:ComparisonDOS}
the density of states (DOS) shifted for the chemical potential $\mu$
of a cluster of 100 water molecules, obtained with our program, and
with the all-electron calculation performed in the commercially available
quantum chemistry program Q-Chem \citep{shao2015advances}. Our code
used $\beta=100E_{h}^{-1}$ but we tested also larger values of $\beta$
to ascertain that the results are visibly identical. We made comparisons
using three different basis sets, ranging from single to triple zeta
quality (STO-3G, 6-31G and 6-311G). To complement the picture, we
also give the frontier orbital energies, band gaps and chemical potentials
corresponding to these calculations in Table~\ref{tab:levels}.

Looking at the shifted DOS, both the results of Q-Chem
and those of the present code converge to indistinguishable values
close to that of the all-electron highest quality basis calculation.
This validates our present code's calculations, even though a small
shift still exists between the chemical potentials (-0.6 eV), as seen
in Table~\ref{tab:levels}. It is noteworthy that the DOS in our
code is less sensitive to basis set quality than the all-electron
code, where for the smallest STO-3G basis set the all-electron calculations
deviate strongly from the converged basis set values, showing a large
($6.3$eV) shift and a band gap which is more than a factor two too
large. The stability of our calculations in comparison to Q-Chem can
be attributed to the use of the norm-conserving pseudopotentials.
Indeed, in the supplementary material F we show that effective core
potentials stabilize the Q-Chem small basis set calculations as well.

An additional validation of our approach can be found
in the supplementary material G, where we compare the potential energy
surface of the $\text{H}_{2}$ molecule calculated with both our code
and Q-Chem and where we show the influence of the grid spacing on
the accuracy of the calculation. Overall, the approximations that
we employ lead to a systematic difference of $\sim0.2\text{\%}$ in
the electronic energy when compared with Q-Chem for most of the examined
distance range (and maximally $\sim0.4\%$) and a small corrugation
which appears when the gridpoint spacing is larger than the width
of the smallest Gaussian primitive. The relative errors in the electronic
energy, and the fact that they are mostly a rigid shift, lead to deviance
of the order of 0.05eV in the bond energy, much smaller than typical
6-311G basis set errors\citep{jensen2017theelephant}.

\begin{table}
\begin{centering}
\begin{tabular}{|>{\raggedright}m{1.5cm}|l|c|c|c|c|}
\hline 
Basis & Method & \small{}$\varepsilon_{H}$ & \small{}$\varepsilon_{L}$ & \small{}$\varepsilon_{g}$ & \small{}$\mu$\tabularnewline
\hline 
\hline 
\multirow{3}{1.5cm}{\small{}STO-3G} & \small{}Present & -4.5 & -2.6 & 1.9 & -3.6\tabularnewline
\cline{2-6} 
 & \small{}All-electron & 1.3 & 4.5 & 3.1 & 2.9\tabularnewline
\cline{2-6} 
 & \small{}SBKJC & -5.3 & -1.6 & 3.7 & -3.4\tabularnewline
\hline 
\multirow{2}{1.5cm}{\small{}6-31G} & \small{}Present & -3.5 & -1.9 & 1.5 & -2.7\tabularnewline
\cline{2-6} 
 & \small{}All-electron & -3.5 & -2.2 & 1.4 & -2.8\tabularnewline
\hline 
\multirow{2}{1.5cm}{\small{}6-311G} & \small{}Present & -4.7 & -3.3 & 1.4 & -4.0\tabularnewline
\cline{2-6} 
 & \small{}All-electron & -4.2 & -2.7 & 1.4 & -3.4\tabularnewline
\hline 
\end{tabular}
\par\end{centering}
\caption{\label{tab:levels}Comparison of frontier energy levels,
the bandgap $\varepsilon_{g}=\varepsilon_{L}-\varepsilon_{H}$ and
the chemical potential $\mu=\left(\varepsilon_{H}+\varepsilon_{L}\right)/2$
for the DOS calculations of Figure~\ref{fig:ComparisonDOS}.}
\end{table}

\section{\label{sec:Scaling-properties-of}Scaling properties of the method }

In this section we study the method's algorithmic complexity and analyze
the speedup achievable by parallelization in terms of strong and weak
scalability.

\subsection{\label{subsec:Algorithmic-complexity}Algorithmic complexity }

\begin{figure*}
\includegraphics[width=0.98\textwidth]{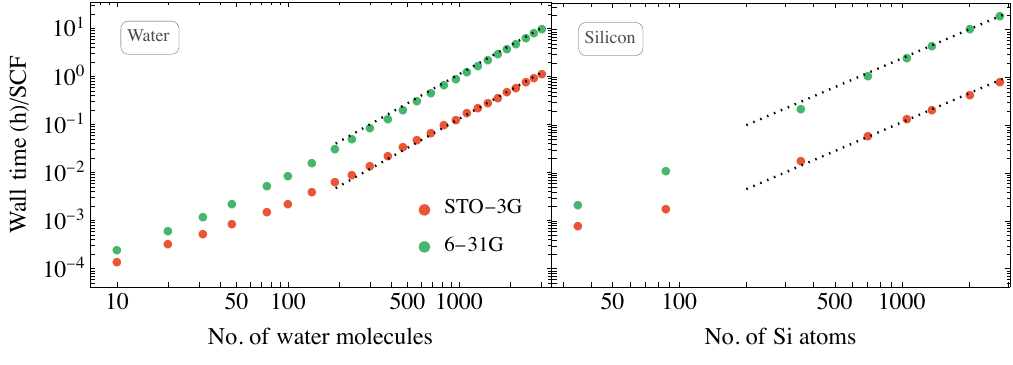}

\caption{\label{fig:QuadraticScaling}The wall-time as a function of system
size, for the water clusters and the silicon nanocrystals, calculated
using two basis sets within the LDA. The calculations were performed
on eight Intel Xeon Gold 6132 CPU @ 2.60GHz 755GB RAM (connected through
Infiniband), using 112 cores for all systems. The dotted lines in
the figures are guides to the eye with designated quadratic scaling.
Fitting the function $t=Ax^{n}$ to the data at the larger time range,
one obtains the exponent $n=1.9\,\left(2.1\right)$ for both the water
and the silicon systems in the STO-3G (6-31G) basis.}
\end{figure*}
To understand the algorithmic complexity of our method, we have to
examine how each part of our code scales as we increase the system
size $K$. Here we are especially interested in the asymptotic behavior,
meaning that the program part with the largest scaling will determine
the overall algorithmic complexity. Our entire SCF cycle, that is
described in detail in the supplementary material A.3, includes different
integral calculations, solving the Poisson equation and calculating
the density. The integral calculation is expected to scale linearly
with system size $K$, i.e. $O(K)$, because the relevant matrices
($F^{KS}$, $S$) are expected to become sparse (see also Figure~\ref{fig:Sparsity}).
The Poisson equation is solved by a fast Fourier transform (FFT) which
scales as $O\left(N_{g}\log N_{g}\right)$, where $N_{g}$ are the
grid points, expected to scale linearly with system size. This leaves
only the density calculation which is done according to equation~\ref{eq:col-by-col}.
The application of the DM $P$ to a column vector $u_{\alpha}$, expressed
through a Chebyshev series, involves repeated applications of the
operator $S^{-1}F^{KS}$ to the column vector $v=S^{-1}u_{\alpha}$
(see supplementary material B for details). The length of the Chebyshev
expansion, $N_{C}$, is independent of the system size $K$ and so
the algorithmic complexity of the $Pu_{\alpha}$ operation is identical
to that of one $S^{-1}F^{KS}v$ operation, namely linear with $K$.
There are a total of $K$ different $Pu_{\alpha}$ operations (see
Eq.~(\ref{eq:col-by-col})), so that the overall algorithmic complexity
of the method is asymptotically quadratic, i.e. $O\left(K^{2}\right)$.
As the system size grows our algorithm could be modified to take advantage
of the emerging sparsity of the DM, allowing for a $K$-independent
complexity of each $Pu_{\alpha}$ operation. In such situations one
can expect an overall linear-scaling numerical complexity, i.e. $O\left(K\right)$.
However, in the present paper, we focus on the broad class of systems
which are very large but for which the DM has not yet localized. Hence
we are in the formally quadratic complexity regime.

To show that quadratic complexity is indeed what we achieve with this
method, we plot, in Figure~\ref{fig:QuadraticScaling} the wall-time
per SCF cycle vs. system size for water clusters (taken from \url{http://www.ergoscf.org/xyz/h2o.php}, accessed on March 5 2022)
and hydrogen-terminated silicon nanocrystals (we use a series of nanocrystals,
starting from \ch{Si35H36} reaching \ch{Si2785H780}, for details,
see supplementary material D), using STO-3G and the larger 6-31G basis
sets. Going from the smaller to the larger basis
set increases wall-time by a factor of 10-20. This result is a combination
of several characteristics beyond the mere size of the basis set.
For example the magnitude of the Gaussian exponents of the basis set's
primitives are relevant for the dimensioning of the grid. Higher valued
Gaussian exponents require a finer mesh and also increase the kinetic
energy component of the Hamiltonian, which increases the Chebyshev
expansion length. Smaller (diffuse) Gaussian exponents lead to larger
grid windows (see also supplementary material A.1.1) and hence an
increase in overall grid size as well. Furthermore, the implementation
of the linear scaling operation of $S^{-1}$, involving the incomplete
Cholesky decomposition and preconditioned conjugate gradients algorithms,
is sensitive to the condition number of $S$, determined by near linear
dependencies between basis functions. As seen in the figure, all
cases show overall quadratic algorithmic complexity. It is noteworthy
to state that the small and intermediate sized systems in the figure
exhibit a varying algorithmic complexity with system size associated
with the interplay between linear complexity processes having a large
prefactor and cubic stages due to the non-sparse nature of the Hamiltonian
and overlap matrices.

\subsection{\label{subsec:Strong-scaling}Strong scalability}

\begin{figure}
\centering{}\includegraphics[width=1\columnwidth]{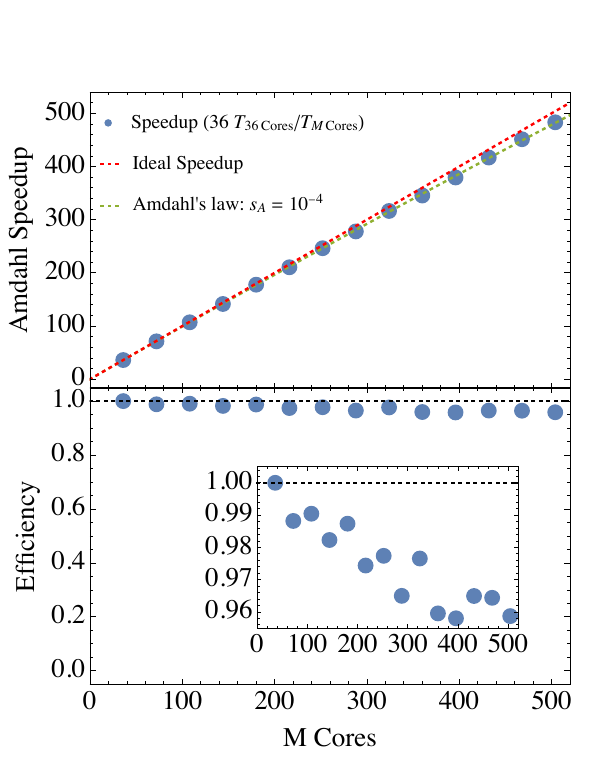}\caption{\label{fig:scalability}Strong scalability speedup analysis (upper
panel) and efficiency $\mathbb{E}\left(M\right)=\mathbb{S}\left(M\right)/M$
(lower panel) for \ch{Si1379H476}. The reference time for 1 processor
for the speedup is extrapolated from $T_{1}=36T_{36}$. The inset
in the lower panel enables a higher resolution of the efficiency regime
close to unity. The calculation used the 6-31G basis set (11984 basis
functions) within the LDA and were performed on several 2.60GHz Intel
Xeon Gold 6240 with 256 GB using 10Gb Ethernet networking communications.}
\end{figure}

In Figure~\ref{fig:scalability} we study the strong scalability
properties of our code, i.e. the scalability achievable when increasing
the number of processors for a \emph{given }task. We show in the figure
the speedup and efficiency for a single SCF iteration of the \ch{Si1379H476}
nanocrystal. Our definition for the speedup in Eq.~\ref{eq:Define-speedup}
requires the knowledge of the elapsed wall time it takes a single
processor (more accurately 1 core) to finish this nanocrystal calculation.
Due to (human) time constraints we had to extrapolate this timing
from a calculation on 36 cores on one single compute node by $T_{1}=36T_{36}$.
The results can be analyzed in terms of the Amdahl law finding that
the the serial fraction is $s_{A}=9\times10^{-5}$ showing a high
degree of parallelization.  Accordingly, the parallelization efficiency
drops very slowly as the number of processors increases, with 96\%
efficiency even at $M=500$ (see the inset in the top panel). We emphasize
that this is achieved with a 10Gb ethernet network communication.
Potentially, the decay of efficiency may be slowed down by employing
a faster communication solution. According to Amdahl's law, efficiency
will drop to $0.5$ when $M\approx\frac{1}{s_{A}}=10^{4}$ . In the
supplementary material E we show results for a smaller system, \ch{Si705H300}
where the Amdahl serial fraction is larger, $s_{A}=2\times10^{-4}$,
a system-size dependency due to the quadratic complexity of our method
(see our discussion in Section~\ref{sec:Introduction}). 

\subsection{\label{subsec:Weak-scaling}Weak scalability }

\begin{figure*}
\centering{}\includegraphics[width=0.44\textwidth]{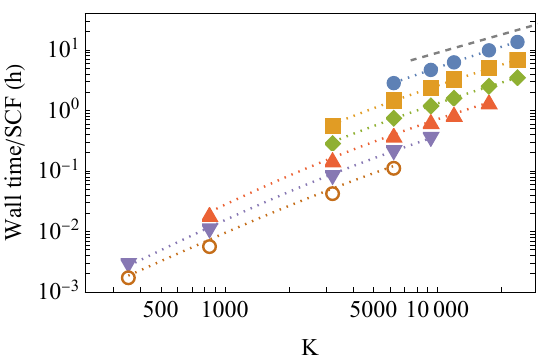}
\includegraphics[width=0.55\textwidth]{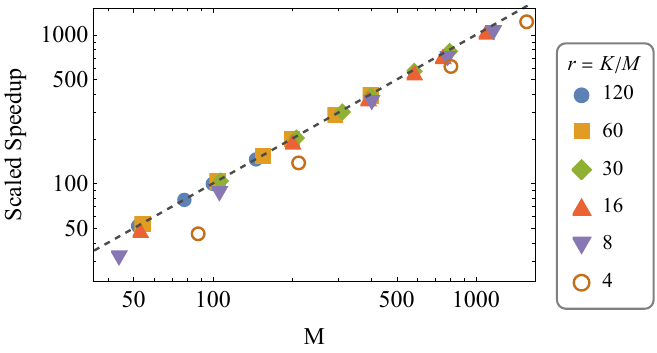}\caption{\label{fig:WeakScaling}Weak scalability speedup analysis. Left: The
wall-time for a single SCF cycle vs. the number of basis set functions
$K$ in calculations given for several fixed values of $r$ ($K/M$
where $M$ is the number of processors) on a series of eight hydrogen-terminated
silicon nanocrystals (detailed in the supplementary material D) using
the 6-31G basis set within the LDA. The black-dashed line is a guide
to the eye showing linear scaling wall-time. The calculations were
run on several Intel Xeon Gold 6240 CPUs @ 2.60GHz 256GB RAM connected
through a 10Gb ethernet networking communications. We had access to
at most $1584$ cores, therefore for $r=4$ we could not treat systems
greater than $K=6420$ and similar though less stringent limitations
appeared for $r=8$ and $16$. The colored dotted lines are the best-fit
of the data to our model in Eq.~\ref{eq:Fit} Right: The scaled speedup
as a function of the number of processors $M$, calculated for the
six values of $r$ from Eq.~\ref{eq:ModelSU} using the best-fit
parameters of our model. The black-dashed line indicates the ``perfect''
speedup $\mathbb{S}=M$.}
\end{figure*}

In this section we focus on the weak scalability properties of our
method, namely how the wall time changes with system size $K$ when
the number of processors afforded to the calculation $M$ grows in
fixed proportion $r=K/M$. In the left panel of Figure~\ref{fig:WeakScaling}
we present the wall-time $T$ as a function of system size $K$ for
six series of runs we made with different fixed ratios ranging from
$r=4$ up to $r=120$ (in the actual calculation, $r$ is the number
of vectors $u_{\alpha}$ assigned to each processor (see Eq.~(\ref{eq:col-by-col}))).
The markers of each series fall on asymptotically straight lines in
the log-log plot which appear parallel to the dark-dashed line indicating
a constant slope of $1$. This confirms the claim of achieving linear-scaling
wall-time in this regime of operation, where $r$ is held constant.
We would also like to examine the \emph{speedup} in order to determine
the degree of efficiency of our calculation on the parallel machine.
For calculating the speedup under our definition in Eq.~(\ref{eq:Define-speedup})
we need to be able to estimate the wall time $T_{1}\left(W\right)$,
which for the large systems is not easily accessible due to (human)
time constraints. Therefore, we developed the following model for
the wall time, with which we will estimate the $M=1$ wall times: 

\begin{equation}
T\left(K,M\right)=\frac{\tau_{2}K^{3}/\left(K_{0}+K\right)}{M}+\tau_{1}K\log M.\label{eq:WallTimeModel}
\end{equation}
The first term on the right is the dominant parallelizable part of
the calculation run on $M$ processors (electron density calculation,
see the supplementary material C for more information). For $K\gg K_{0}$
it exhibits quadratic scaling while for $K\ll K_{0}$ the scaling
is cubic due to insufficient sparsity of the Hamiltonian and overlap
matrices for small $K$. The second term in Eq.~(\ref{eq:WallTimeModel})
reflects the timing of the serial part of the calculation, dominated
by the communication time needed for specific MPI functions (reduce
and broadcast) and scales linearly with $K$ and logarithmically with
$M$. 

\textit{\emph{Using the analytical model, the speedup can now be }}obtained
by plugging Eq.~(\ref{eq:WallTimeModel}) into Eq.~(\ref{eq:Define-speedup}),
resulting in the following closed form expression, 
\begin{equation}
\mathbb{S}\left(r,M\right)=\frac{M}{\left(\frac{K_{0}}{rM}+1\right)\frac{\tau_{1}}{\tau_{2}r}\log M+1}.\label{eq:ModelSU}
\end{equation}
From this equation, it can be seen, that for asymptotically large
values of $M$, the speedup approaches the limit $\mathbb{S}\left(r,M\right)\to\frac{M}{\frac{\tau_{1}}{\tau_{2}r}\log M+1}$
and as long as $r$ is not too small, 
\begin{equation}
r>\frac{\tau_{1}}{\tau_{2}}\log M,\label{eq:condition on r}
\end{equation}
 the speedup is close to ideal $\mathbb{S}\left(r,M\right)\to M$. 

We now fit our model to the calculation's timing results from the
six constant-$r$ series shown in the left panel of Figure~\ref{fig:WeakScaling}
(a total of 32 data points). This leads to a best-fit set of parameters
(in hours): $\tau_{1}\to5.16\times10^{-6}\,\text{h}$ and $\tau_{2}\to5.63\times10^{-7}\text{h}$
and $K_{0}\to2292.6$ for our model and the resulting fit functions,
\begin{equation}
\frac{T_{\text{fit}}\left(K;r\right)}{10^{-6}\text{h}}=K\left(\frac{0.563 r}{\frac{2292.6}{K}+1}+5.16 \log\left(K/r\right)\right)\label{eq:Fit}
\end{equation}
are plotted in the left panel of the figure as dotted colored lines,
one for each values of $r$. It can be seen that these fit functions
indeed reproduce the actual data (given as points) quite closely. 

Having the best-fit parameters, let us now discuss the actual estimated
values for the (scaled) speedup in the Gustafson sense. These estimates,
based on Eq.~(\ref{eq:ModelSU}) are plotted in the right panel of
Figure~\ref{fig:WeakScaling}. We see that for $r>16$ the the speedup
is not too far from ideal, in accordance with the analysis presented
above, however as indicated in Eq.~(\ref{eq:condition on r}) the
speedup is smaller when $r$ decreases as is clearly visible for $r=8$
and small $M$ and for $r=4$ for all values of $M$. However, even
for these small $r$ cases, the speedup is maintained as $M$ increases
and the calculation is still quite efficient.

\section{\label{sec:Summary-and-Conclusions}Summary and Conclusions}

In this paper we presented a parallelizable electronic structure approach
to finite temperature density functional theory under (semi)local
functionals, using atom-centered Gaussian basis sets which offers
linear wall-time complexity as a function of system size in the weak
scalability regime. The inherent time complexity of the method is
quadratic $O\left(K^{2}\right)$, as discussed in section~\ref{subsec:Algorithmic-complexity}
and it does not involve truncation of density matrix elements, characteristics
of linear-scaling approaches. 

Our trace-based calculation combined with Chebyshev expansions allows
for efficient parallelization in the strong scalability sense, as
shown in subsection~\ref{subsec:Strong-scaling}. Due to the quadratic
complexity, we found that the value of the Amdahl parameter was system-size
dependent, with $s_{A}=2\times10^{-4}$ for the \ch{Si705H300} system
and $s_{A}=9\times10^{-5}$ for \ch{Si1379H476}. The overall weak
scalability performance shows that linear scaling wall time is achievable,
as demonstrated in section~\ref{subsec:Weak-scaling} and is highly
efficient when the number of orbitals per processors $r$ is not smaller
than $\sim$10 and beyond that efficiency drops by a factor of $\sim$1.5. 

Our main conclusion is, that this type of approach has the potential
to be a useful and efficient tool for studying large systems in regimes
where quantum confinement or electron delocalization prevents traditional
linear-scaling to set in. Furthermore, for even larger systems, where
electrons localize, we plan to enable linear scaling either through
stochastic orbital methods \citep{fabian2019stochastic} or by exploiting
directly the DM's finite range. While in this paper
we were concerned mainly with the scalability of the density calculation,
force evaluation, done after the density converges, is also an important
goal, high on our list of future plans. We will follow our recent
work developing a stochastic estimation of the exact energy derivative
(Hellman-Feynman) forces \citep{shpiro2021forcesfrom}. As shown in
ref.~\citep{arnon2017equilibrium}, these stochastic estimations
lead to noisy forces that can be used only within Langevin dynamics.
In our case, we expect that the deterministic evaluation of the exact
derivatives will result in deterministic forces of sufficient quality
to enable energy-conserving molecular dynamics simulations. 

\section*{Acknowledgments}

We gratefully acknowledge support from the Israel Science Foundation
grant 800/19. MF expresses special thanks to Roie
Dann for helpful discussions.

\section*{Data availability}

Binaries that support the findings of
this study are available from the corresponding author upon reasonable
request.

\section*{Supporting Information}
Details on our implementation of the Gaussian integral evaluation on the Cartesian grid, the Chebychev expansion, our parallelization strategy, systems studied, density of states calculated with Q-Chem and effective core potentials and the $\textnormal{H}_{2}$ potential energy surface.
This information is available free of charge via the Internet at http://pubs.acs.org

\bibliography{QuadraticPaper1}

\end{document}


\appendix

\section{\label{sec:Basis-set---grid}basis set / grid representation}

\subsection{\label{subsec:Basis2Grid}basis set $\to$ grid transformation}

Each basis function $\phi_{\alpha}\left(\mathbf{r}\right)$ is centered
on an atom, located at point $\boldsymbol{R_{\alpha}}$ and associated
with $l_{\alpha}$, and $m_{\alpha}$, the angular momentum and magnetic
quantum number, respectively. The basis functions employed are non-orthogonal
Gaussians that are contracted to Gaussian type orbitals (GTOs):
\begin{equation}
\phi_{\alpha}\left(\mathbf{r}\right)=\sum_{p=1}^{N_{cn}}d_{\alpha}^{p}\Phi\left(\mathbf{r}-\boldsymbol{R}_{\alpha};\zeta_{\alpha}^{p},l_{\alpha},m_{\alpha}\right),\label{eq:Contraction-1}
\end{equation}
where $\zeta_{\alpha}^{p}$ are the Gaussian exponents, $d_{\alpha}^{p}$
are the contraction coefficients $N_{cn}$ is the contraction length,
and 
\begin{equation}
\Phi\left(\mathbf{r};\zeta,l,m\right)=\sum_{i,j,k}w_{lm}^{i,j,k}\varphi^{ijk}\left(\boldsymbol{r};\zeta\right)\label{eq:SphericalHarmonicTransformation}
\end{equation}
are the primitive spherical GTOs which are constructed from the following
Cartesian GTOs 
\begin{equation}
\varphi^{ijk}\left(\boldsymbol{r};\zeta\right)=x^{i}e^{-\zeta\:x^{2}}\times y^{j}e^{-\zeta\:y^{2}}\times z^{k}e^{-\zeta\:z^{2}},\label{eq:primitiveCartesianGaussian}
\end{equation}
where $\boldsymbol{r}=\left(x,y,z\right)$, $i+j+k=l$ and the weights
$w^{ijk}$ are given in Table~\ref{tab:SphericalHarmonicGTO}. 

In order to evaluate the GTO primitive at a grid-point $\boldsymbol{r}_{g}=\left(x_{g},y_{g},z_{g}\right)$,
for each Cartesian dimension, e.g. $x$, we store in memory an array
containing, the values of $\left(x_{g}-R_{\alpha x}\right)^{i}e^{-\zeta_{\alpha_{p}}\:\left(x_{g}-R_{\alpha x}\right)^{2}}$
(and its derivatives) in the interval $\left[-L_{\alpha x},L_{\alpha x}\right]$
around $R_{\alpha x}$. The memory cost for the three arrays is small
since $L_{\alpha}$ is small due to the compact nature of the Gaussian
function. With these considerations, the evaluation of $\varphi^{ijk}\left(\mathbf{r}_{g},\zeta\right)$
involves just two multiplications, at least an order of magnitude
faster than a direct evaluation of the GTO primitive.

\begin{table*}
\centering{}\caption{Cartesian angular momenta and weight derived from the real solid harmonics\citep{helgaker2014molecular}
for $l\protect\leq2$. Note that only cases where $i+j+k=l$ give
non-zero values for $d$ and $w$.}
\label{tab:SphericalHarmonicGTO}\setlength{\tabcolsep}{0.7em}
{\global\long\def\arraystretch{1.5}
\begin{tabular}{|c|c|c|c|c|cc|c|ccc|c|c|}
\hline 
\textbf{$l$} & $s$ & \multicolumn{3}{c|}{$p$} & \multicolumn{8}{c|}{$d$}\tabularnewline
\hline 
$m$ & $0$ & $1$ & $0$ & $-1$ & \multicolumn{2}{c|}{$2$} & $1$ & \multicolumn{3}{c|}{$0$} & $-1$ & $-2$\tabularnewline
\hline 
$i$ & $0$ & $1$ & $0$ & $0$ & $2$ & $0$ & $1$ & $2$ & $0$ & $0$ & $0$ & $1$\tabularnewline
\hline 
$j$ & $0$ & $0$ & $0$ & $1$ & $0$ & $2$ & $1$ & $0$ & $2$ & $0$ & $1$ & $0$\tabularnewline
\hline 
$k$ & $0$ & $0$ & $1$ & $0$ & $0$ & $0$ & $0$ & $0$ & $0$ & $2$ & $1$ & $1$\tabularnewline
\hline 
$w^{ijk}$ & $1$ & $1$ & $1$ & $1$ & $\nicefrac{\sqrt{3}}{2}$ & $-\nicefrac{\sqrt{3}}{2}$ & $\sqrt{3}$ & $-\nicefrac{1}{2}$ & $-\nicefrac{1}{2}$ & $1$ & $\sqrt{3}$ & $\sqrt{3}$\tabularnewline
\hline 
\end{tabular}}
\end{table*}

\subsubsection{\label{subsubsec:The-GTO-window}The GTO window }

\begin{figure}[htbp]
\includegraphics{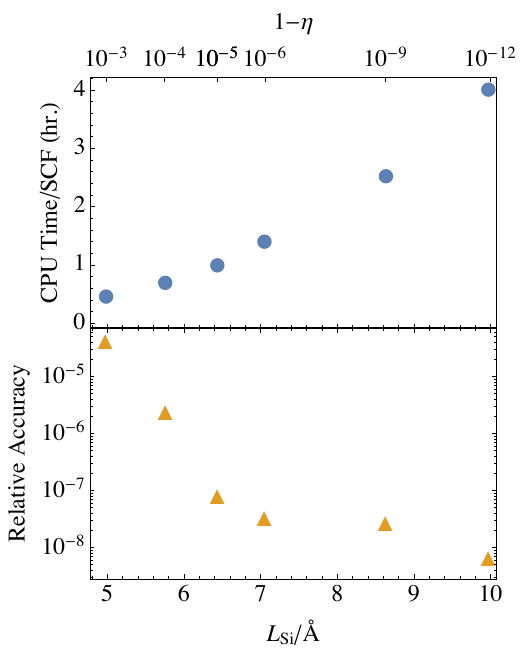} \caption{\label{fig:cutoff}The wall-time of a deterministic LDA calculation
(cubic-scaling) of Si$_{353}$H$_{196}$ and the accuracy in the relative
energy vs. $\eta$ and the silicon atom widow length $L_{Si}$. The
grid spacing was $\Delta h=0.5a_{0}$ and the basis set was 6-31G.}
\end{figure}

The length $L_{\alpha}$ of the window encompassing the basis function
$\phi_{\alpha}\left(\boldsymbol{r}\right)$ is determined such that
the most protruding primitive GTO has $1-\eta$ of its ``charge''
included, where $\eta$ is a small cutoff parameter. This can be expressed
as the solution of the charge equation 
\begin{equation}
c\left(L_{\alpha}\sqrt{\zeta_{\alpha}},l_{\alpha}\right)=1-\eta,\label{WindowLength}
\end{equation}
where $c\left(d,l\right)\equiv\frac{\int_{0}^{d}x^{l}e^{-x^{2}}dx}{\int_{0}^{\infty}x^{l}e^{-x^{2}}dx}$
and $\zeta_{\alpha}=\min_{p}\zeta_{\alpha}^{p}$ identify the exponent
of the most protruding primitive GTO. An important issue is the choice
of $\eta$, small values give larger windows, improving accuracy at
the expense of a higher computational cost. In Figure~\ref{fig:cutoff}
the wall-time and accuracy is plotted vs the $\eta$ parameter and
the window length $L$ for a deterministic calculation of Si$_{353}$H$_{196}$.
As $\eta$ drops from $10^{-3}$ to $10^{-6}$, $L$ grows from $5$
to $7~\text{\AA}$ (the volume and number of grid-points grow by a factor
of 2.7), the single SCF iteration wall-time increases by a factor
of 3, while the relative energy error drops by nearly 3 orders of
magnitude. 

\begin{figure}[htbp]
\centering{}\includegraphics{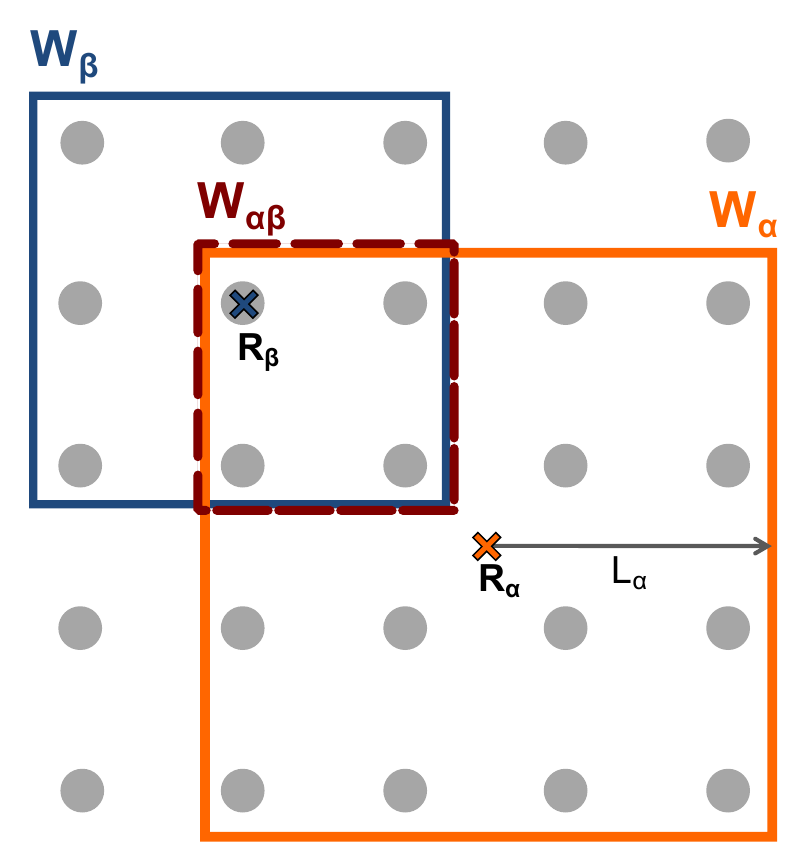} \caption{\label{fig:Windows}Schematic of grid windows in 2D. Shown are the
grid points displayed as grey disks and two atomic centers, R$_{\alpha}$
and R$_{\beta}$, with their respective GTO windows, W$_{\alpha}$
and W$_{\beta}$ of length L$_{\alpha}$ and L$_{\beta}$(not shown).
The window W$_{\alpha\beta}$ is the region of overlap between W$_{\alpha}$
and W$_{\beta}$.}
\end{figure}

Furthermore, not all grid points of the entire grid will be relevant
to evaluate the numerical integrals for each pair of basis function.
We therefore introduce for each contracted spherical GTO $\phi_{\alpha}\left(\mathbf{r}_{g}\right)$
a small grid-axes aligned cubic window $W_{\alpha}$ of lengths $\boldsymbol{L}_{\alpha}=\left(L_{\alpha x},L_{\alpha y},L_{\alpha z}\right)$,
in which the basis function is defined and nonzero. In Figure~\ref{fig:Windows}
two such grid windows are displayed for the 2 dimensional case for
illustrative purpose (our program uses of course 3D windows): two
atomic centers are present (R$_{\alpha}$, R$_{\beta}$), each with
their own window (W$_{\alpha}$, W$_{\beta}$). For each of the windows
only a subset of points is relevant and even more so for the overlap
of both windows, W$_{\alpha\beta}$, where only four grid points are
relevant in this example.

\subsubsection{\label{subsubsec:n(r)}The density $n\left(\boldsymbol{r}_{g}\right)$}

One important step in the DFT algorithm is to evaluate the electronic
density on the grid, based on the density matrix. In order to calculate
the electronic density, we define
\begin{equation}
n_{\alpha}\left(\mathbf{r}_{g}\right)=2\phi_{\alpha}\left(\mathbf{r}_{g}\right)\sum_{\beta}\phi_{\beta}\left(\mathbf{r}_{g}\right)P_{\beta\alpha}\label{eq:ElectronDensity}
\end{equation}
as the density contributed by the $\alpha$'s column of the density
matrix $P$, which can be calculated independently on parallel processors.
The calculation is done by the following schematic algorithm:
\begin{enumerate}
\item For all $\boldsymbol{r}_{g}\in W_{\alpha}$, set $n_{\alpha}\left(\boldsymbol{r}_{g}\right)=0$.
\item For each $\beta$ such that $W_{\alpha\beta}=W_{\alpha}\cap W_{\beta}$
(see Figure~\ref{fig:Windows}) and $\boldsymbol{r}_{g}\in W_{\alpha\beta}$,
we update the density $n_{\alpha}\left(\boldsymbol{r}_{g}\right)\leftarrow n_{\alpha}\left(\boldsymbol{r}_{g}\right)+\phi_{\alpha}\left(\mathbf{r}_{g}\right)\phi_{\beta}\left(\mathbf{r}_{g}\right)P_{\beta\alpha}$.
\end{enumerate}
The total density $n\left(\mathbf{r}\right)=\sum_{\alpha}n_{\alpha}\left(\boldsymbol{r}\right)$
is then obtained as a reduce operation for all processors. For more
details on the parallelization see section \ref{sec:Parallelization}.

\subsection{\label{subsec:Grid2Basis}Grid $\to$ basis set transformation}

Here we will explain how to construct the AO matrices from the corresponding
grid-operators, the KS potential $v_{KS}\left(\mathbf{r}\right)\to V_{\alpha\beta}^{KS}$,
the non-local pseudopotential $\hat{v}_{nl}\left(\mathbf{r},\mathbf{r}'\right)\to V_{\alpha\beta}^{\text{NL}}$
and the kinetic energy $-\frac{\hbar^{2}}{2m_{e}}\nabla^{2}\to T_{\alpha\beta}$.

\subsubsection{\label{subsubsec:V^KS}KS potential $V^{KS}$ and overlap $S$}

In order to generate a matrix element $V_{\alpha\beta}^{KS}$ in the
basis set representation from a grid-based function $v_{KS}\left(\mathbf{r}\right)$,
we only have to sum over the grid points, where both windows $W_{\alpha}$
and $W_{\beta}$ overlap ($W_{\alpha\beta}=W_{\alpha}\cap W_{\beta}$,
see also Figure~\ref{fig:Windows}) : 
\begin{equation}
V_{\alpha\beta}^{KS}=\negmedspace\sum_{\mathbf{r}_{g}\in W_{\alpha\beta}}\negmedspace\phi_{\alpha}\left(\mathbf{r}_{g}\right)v_{KS}\left(\mathbf{r}_{g}\right)\phi_{\beta}\left(\mathbf{r}_{g}\right)h^{3},\label{eq:Grid2BS}
\end{equation}
with $h$ as the grid spacing. A list is kept in advance for tracking
the pairs of orbitals $\phi_{\alpha}$ and $\phi_{\beta}$ that actually
have a non empty $W_{\alpha\beta}$, expediting the evaluation. The
overlap matrix can be calculated this way taking $v_{s}\left(\mathbf{r}_{g}\right)=1$
at each gridpoint.

\subsubsection{\label{subsubsec:V^PP}Non-local pseudopotential $V^{NL}$}

While the grid enables a very efficient evaluation of the KS-potential,
it is not ideal for describing core electrons. Therefore these electrons
are taking into account through norm-conserving pseudopotentials.
For atom $C$ at the location $\mathbf{R}_{C}$, there is, similar
to the window for a basis function, a window $W^{C}$, i.e. a set
of grid points for which the nonlocal part of the pseudopotential,
$v_{nl}^{\text{C}}\left(\mathbf{r}_{g}-\mathbf{R}_{C},\mathbf{r}_{g}^{\prime}-\mathbf{\mathbf{R}}_{C}\right)$
operates. The matrix $V^{\text{NL }}$ representing the non-local
part of the pseudopotential in the AO basis has non zero elements
$V_{\alpha\beta}^{\text{NL }}$ only for those orbital pairs for which
there exists an atom $C$ for which $W_{\alpha}^{C}\equiv W^{C}\cap W_{\alpha}\ne\emptyset$
and $W_{\beta}^{C}\ne\emptyset$ (where $\emptyset$ is the empty
set of grid points). In practice, the calculation of $V^{\text{NL}}$
is done by locating, for each $\beta$ all atoms $C$ for which $W_{\beta}^{C}\ne\emptyset$
and then calculating the localized function:
\begin{equation}
\psi_{C\beta}\left(\mathbf{r}_{g}\right)=\negmedspace\sum_{\boldsymbol{r}_{g}^{\prime}\in W_{\beta}^{C}}\negmedspace\negmedspace v_{nl}^{\text{C}}\left(\mathbf{r}_{g}-\mathbf{R}_{C},\mathbf{r}_{g}^{\prime}-\mathbf{\mathbf{R}}_{C}\right)\phi_{\beta}\left(\mathbf{r}_{g}^{\prime}\right)h^{3},\quad\mathbf{r}_{g}\in W^{C}.
\end{equation}
Then one searches for all orbitals $\phi_{\alpha}$ for which $W_{\alpha}^{C}\ne\emptyset$
and calculates 
\begin{equation}
V_{\alpha\beta}^{C}=\negmedspace\sum_{\boldsymbol{r}_{g}\in W_{\alpha}^{C}}\negmedspace\negmedspace\phi_{\alpha}\left(\mathbf{r}_{g}\right)\psi_{C\beta}\left(\mathbf{r}_{g}\right)h^{3}
\end{equation}
finally, the matrix element is obtained as a sum over all atoms $C$
for which $W_{\beta}^{C}\ne\emptyset$:

\begin{equation}
V_{\alpha\beta}^{\text{NL}}=\negmedspace\sum_{C:W_{\beta}^{C}\ne\emptyset}\negmedspace V_{\alpha\beta}^{C}.\label{eq:PP2BS}
\end{equation}
The pseudo-potential matrix $V^{\text{NL }}$ has to be calculated
only once in the beginning of the calculation.

\subsubsection{\label{subsubsec:T}Kinetic energy $T$}

The final type of integral considered here is the kinetic energy integral.
\begin{equation}
T_{\alpha\beta}=\frac{1}{2}\sum_{\mathbf{r}_{g}\in W_{\alpha\beta}}\negmedspace\negmedspace\negmedspace\left[\nabla\phi_{\alpha}\left(\mathbf{r}_{g}\right)\right]\cdot\left[\nabla\phi_{\beta}\left(\mathbf{r}_{g}\right)\right]h^{3}.\label{eq:T2BS}
\end{equation}
Here we need to take the gradient of the basis function $\phi_{\alpha}$
and $\phi_{\beta}$ and multiply them. Taking the gradient of both
basis functions rather than the Laplace operator for only one ensures
that T will also numerically be positive. For the gradient we need
the partial derivatives of the basis functions calculated at each
grid point $\mathbf{r_{g}}$. Currently derivatives up to second order
are implemented. They are taken with respect to equation~\ref{eq:primitiveCartesianGaussian}.

\subsection{\label{subsec:SCF}Putting everything together: The SCF cycle}

\begin{figure}
\includegraphics{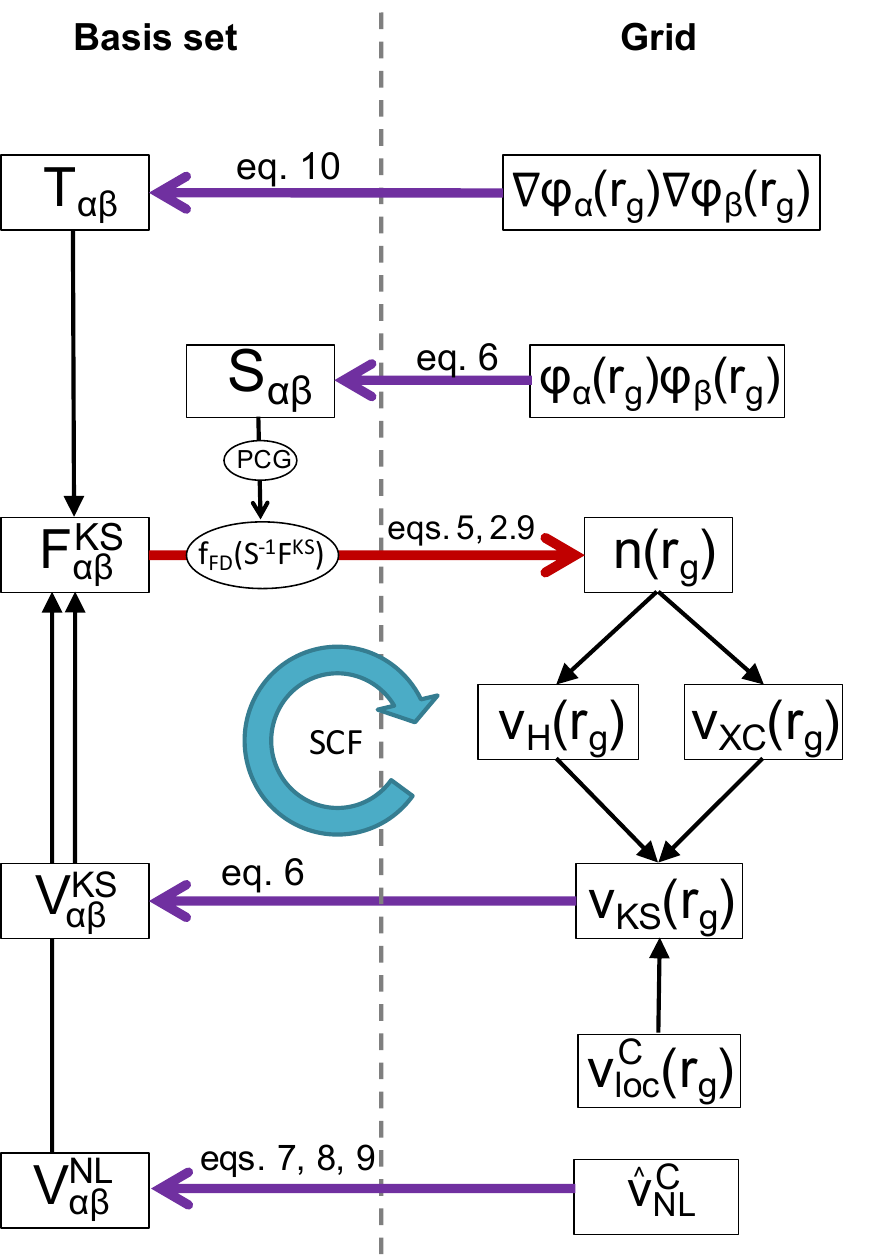}

\centering\caption{\label{fig:Flowchart-KSDFT}Flowchart for the KS-DFT calculation indicating
the stages of the processing and the transformation of data back and
forth between the grid and basis set representation during the SCF
procedure }
\end{figure}

In order to solve the non-linear KS equation, one has to do so iteratively,
until the SCF solution is found. This process is depicted in Figure~\ref{fig:Flowchart-KSDFT},
with a focus on our unique basis set/grid representation. All integrals
are calculated from their respective grid operators, as discussed
in section~\ref{subsubsec:V^KS}-\ref{subsubsec:T}. This grid$\to$basis
set transformation is depicted in Figure~\ref{fig:Flowchart-KSDFT}
by purple arrows and the corresponding equation number. Except for
$V^{KS}$, all other matrices are only calculated once in the beginning
of the SCF cycle. For the start of the SCF cycle the KS-Fock matrix,
$F^{KS}$, is calculated by summing $F^{KS}=T+V^{NL}+V^{KS}$, where
the first $V^{KS}$ matrix is determined by an initial guess of the
density $n\left(\mathbf{r}\right)$ (for us usually the sum of atomic
densities). The next step in the SCF cycle is to update the density
$n\left(\mathbf{r}\right)$ on the grid. This step is the only basis
set$\to$grid transformation and is shown in Figure~\ref{fig:Flowchart-KSDFT}
by a red arrow. As discussed in section~\ref{subsubsec:n(r)}, the
density matrix $P$ is put column-wise on the grid. Each column of
$P$ is calculated according to $P=f_{FD}\left(S^{-1}F^{KS}\right)S^{-1}$,
where the application of the inverse overlap matrix $S^{-1}$ is handled
by a preconditioned conjugate gradient (PCG) approach \footnote{We used the incomplete Cholesky preconditioning \citep{scott2014hslmi28}
for the conjugated gradient approach implemented in the HSL-MI28 and
MI21 codes, respectively, where HSL is a collection of FORTRAN codes
for large scale scientific computation ( http://www.hsl.rl.ac.uk/).}, thereby only requiring the repeated application of $S$ (which is
advantageous, since $S$ is much sparser than $S^{-1}$) and the matrix
exponential ($\exp\left(S^{-1}F^{KS}\right)$) is evaluated by a Chebyshev
expansion (see section~\ref{sec:ChebychevExpansion}). Once the density
is calculated, the Hartree potential $v_{H}\left(\mathbf{r}\right)$
is obtained by solving Poisson's equation through a fast Fourier transform
(FFT) \citep{martyna1999areciprocal} and the exchange-correlation
potential $v_{XC}\left(\mathbf{r}\right)$ by employing the relevant
functional form (in our case the local-density approximation (LDA)).
The KS potential on the grid $v_{KS}\left(\mathbf{r}\right)$ is then
calculated according to $v_{KS}\left(\boldsymbol{r_{g}}\right)=\sum_{C\in nuclei}v_{loc}^{C}\left(\boldsymbol{r_{g}}-\boldsymbol{R}_{C}\right)+v_{H}\left[n\right]\left(\boldsymbol{r_{g}}\right)+v_{xc}\left[n\right]\left(\boldsymbol{r_{g}}\right),$where
$v_{loc}$ is only calculated once in the beginning of the SCF cycle.
The KS potential matrix $V^{KS}$ is updated according to eq.~\ref{eq:Grid2BS}
before the next SCF cycle starts with a new KS Fock matrix.

\section{\label{sec:ChebychevExpansion}Chebyshev expansion}

The inverse temperature parameter $\beta$ in the Fermi-Dirac function
$f_{FD}\left(\varepsilon\right)=\frac{1}{1+e^{\beta\left(\varepsilon-\mu\right)}}$
needs to be chosen high enough so that $\beta\left(\varepsilon_{L}-\varepsilon_{H}\right)\gg1$
where $\varepsilon_{L}$ ($\varepsilon_{H}$) is the Kohn-Sham eigenvalue
of the lowest unoccupied (highest occupied) molecular orbital. The
chemical potential parameter $\mu$ must be adjusted to reproduce
the number of electrons in the system. For the actual application
of the Fermi-Dirac function on a vector $u_{\alpha}$ we need to evaluate
a matrix exponential of a matrix product ($\exp\left(S^{-1}F^{KS}\right)$)
and this is achieved by using the Chebyshev polynomial expansion of
the Fermi-Dirac function
\begin{equation}
f_{FD}\left(\varepsilon\right)=\sum_{l=0}^{N_{C}-1}a_{l}\left(T,\mu\right)T_{l}\left(\frac{H-\bar{E}}{\Delta E}\right)
\end{equation}
where $T_{l}\left(x\right)$ are the Chebyshev polynomials obeying
the recursion relation $T_{l+1}\left(x\right)=2xT_{l}\left(x\right)-T_{l-1}\left(x\right)$,
$N_{C}\approx2\Delta E\beta$ is the expansion length, and finally:
$\Delta E=\frac{E_{max}-E_{min}}{2}$, $\bar{E}=\frac{E_{max}+E_{min}}{2}$,
where $E_{max}$ ($E_{min}$) is the largest (smallest) eigenvalue
of $H=S^{-1}F^{KS}$. The Chebyshev expansion coefficients $a_{l}\left(T,\mu\right)$
can be computed using fast Fourier transform methods \citep{baer1997chebyshev}.
The expansion, when applied to the Fermi-Dirac function of the Hamiltonian
$H$ allows us to approximate the $\alpha's$ column of the DM $\rho_{\alpha}\equiv Pu_{\alpha}$
by the following procedure: First, we set $U_{0}\leftarrow S^{-1}u_{\alpha}$,
$U_{1}\leftarrow H_{N}U_{0}$ (where $H_{N}=\frac{H-\bar{E}}{\Delta E}$
is the ``normalized'' Hamiltonian) and $l=2,$ then we compute $\rho_{\alpha}\leftarrow a_{0}\left(T,\mu\right)U_{0}+a_{1}\left(T,\mu\right)U_{1}$,
and loop until $l=N_{C}$ the following steps:

\begin{align}
U_{2} & \leftarrow2H_{N}U_{1}-U_{0},\label{eq:ChebychevExpansion}\\
\rho_{\alpha} & \leftarrow\rho_{\alpha}+a_{l}(T,\mu)U_{2},\\
U_{0} & \leftarrow U_{1},\,\,\,U_{1}\leftarrow U_{2}\\
l & \leftarrow l+1
\end{align}

\section{\label{sec:Parallelization}Parallelization}

\begin{figure}[htbp]
\centering{}\includegraphics{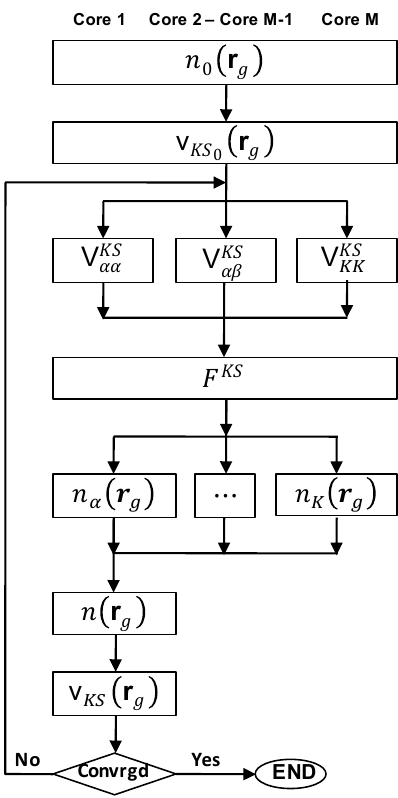} \caption{Parallelization strategy for the SCF cycle employed in the program}
\label{fig:Parallelization} 
\end{figure}

In this section we will discuss the main advantage of this algorithm,
which is the straightforward path to parallelization. To understand
the parallelization strategy a flowchart of the program is depicted
in Figure~\ref{fig:Parallelization}. All parallelization shown here
is done through the \textit{message passing interface} (MPI).

There are a total number of $M$ processors that can be used and between
which the workload in the SCF cycle is split. At first all processors
are producing the same initial guess density and an initial KS potential,
$n_{0}(\textbf{r}_{g})$ and $v_{KS_{0}}(\textbf{r}_{g})$ respectively.
This $v_{KS_{0}}(\textbf{r}_{g})$ is then transferred from the grid
to the matrix representation with equation~\ref{eq:Grid2BS}. This
step is parallelized in such a way that each processor gets a batch
of matrix elements $V_{\alpha\beta}^{KS}$ that it needs to calculate
and in the end these separate matrix elements are combined and distributed
to every processor. The KS-Fock matrix is updated next for every processor,
where $F^{KS}=T+V^{NL}+V^{KS}$. $T$ and $V^{NL}$ are calculated
once before the SCF cycle according to equation~\ref{eq:PP2BS} and~\ref{eq:T2BS}
and are parallelized in the same way as described for $V^{KS}$. Each
processor is independently calculating an electron density on the
grid according to equation~\ref{eq:ElectronDensity}. For this, each
processor gets a different set of unit vectors $u_{\alpha}$, where
the total number of vectors is just the basis set size $K$. Each
processor therefore gets to calculate a fraction $r=\nicefrac{K}{M}$
of the densities. The maximal parallelization that can be achieved
in this step is therefore for $M=K$ and $r=1$. All the densities
from the different processors are summed on the first (``master'')
processor. Only the master process calculates the new potential $v_{KS}(\textbf{r}_{g})$
and checks if the calculation reached the defined convergence criteria
($\Delta E=E^{SCF}-E^{SCF-1}$). If this is the case the calculation
finishes, otherwise $v_{KS}(\textbf{r}_{g})$ is distributed to all
processors and the SCF cycle is repeated from the calculation of the
KS potential matrix. For more details on the SCF cycle see also subsection~\ref{subsec:SCF}.

\section{\label{sec:Systems}Systems}

\begin{table*}
\centering{}\caption{Silicon clusters employed in the calculations}
\label{tab:Silicon_Systems}

\begin{tabular}{|c|c|c|c|c|}
\hline 
System & K (STO-3G) & K (6-31G) & \#$e^{-}$ & \# atoms\tabularnewline
\hline 
\hline 
\ch{Si35H36} & 176 & 352 & 176 & 71\tabularnewline
\hline 
\ch{Si87H76} & 424 & 848 & 424 & 163\tabularnewline
\hline 
\ch{Si353H196} & 1608 & 3216 & 1608 & 549\tabularnewline
\hline 
\ch{Si705H300} & 3120 & 6240 & 3120 & 1005\tabularnewline
\hline 
\ch{Si1063H412} & 4664 & 9328 & 4664 & 1475\tabularnewline
\hline 
\ch{Si1379H476} & 5992 & 11984 & 5992 & 1855\tabularnewline
\hline 
\ch{Si2031H628} & 8752 & 17504 & 8752 & 2659\tabularnewline
\hline 
\ch{Si2785H780} & 11920 & 23840 & 11920 & 3565\tabularnewline
\hline 
\end{tabular}
\end{table*}
\begin{table*}
\centering{}\caption{Water clusters employed in the calculations (taken from \protect\url{http://www.ergoscf.org/xyz/h2o.php}) }
\label{tab:Water_Systems}

\begin{tabular}{|c|c|c|c|c|}
\hline 
System & K (STO-3G) & K (6-31G) & \#$e^{-}$ & \# atoms\tabularnewline
\hline 
\ch{(H2O)10} & 60 & 120 & 80 & 30\tabularnewline
\hline 
\ch{(H2O)20} & 120 & 240 & 160 & 60\tabularnewline
\hline 
\ch{(H2O)32} & 192 & 384 & 256 & 96\tabularnewline
\hline 
\ch{(H2O)47} & 282 & 564 & 376 & 141\tabularnewline
\hline 
\ch{(H2O)76} & 456 & 912 & 608 & 228\tabularnewline
\hline 
\ch{(H2O)100} & 600 & 1200 & 800 & 300\tabularnewline
\hline 
\ch{(H2O)139} & 834 & 1668 & 1112 & 417\tabularnewline
\hline 
\ch{(H2O)190} & 1140 & 2280 & 1520 & 570\tabularnewline
\hline 
\ch{(H2O)237} & 1422 & 2844 & 1896 & 711\tabularnewline
\hline 
\ch{(H2O)301} & 1806 & 3612 & 2408 & 903\tabularnewline
\hline 
\ch{(H2O)384} & 2304 & 4608 & 3072 & 1152\tabularnewline
\hline 
\ch{(H2O)471} & 2826 & 5652 & 3768 & 1413\tabularnewline
\hline 
\ch{(H2O)573} & 3438 & 6876 & 4584 & 1719\tabularnewline
\hline 
\ch{(H2O)692} & 4152 & 8304 & 5536 & 2076\tabularnewline
\hline 
\ch{(H2O)816} & 4896 & 9792 & 6528 & 2448\tabularnewline
\hline 
\ch{(H2O)964} & 5784 & 11568 & 7712 & 2892\tabularnewline
\hline 
\ch{(H2O)11120} & 6720 & 13440 & 8960 & 3360\tabularnewline
\hline 
\ch{(H2O)1293} & 7758 & 15516 & 10344 & 3879\tabularnewline
\hline 
\ch{(H2O)1481} & 8886 & 17772 & 11848 & 4443\tabularnewline
\hline 
\ch{(H2O)1698} & 10188 & 20376 & 13584 & 5094\tabularnewline
\hline 
\ch{(H2O)1924} & 11544 & 23088 & 15392 & 5772\tabularnewline
\hline 
\ch{(H2O)2165} & 12990 & 25980 & 17320 & 6492\tabularnewline
\hline 
\ch{(H2O)2469} & 14814 & 29628 & 19752 & 7407\tabularnewline
\hline 
\ch{(H2O)2737} & 16422 & 32844 & 21896 & 8211\tabularnewline
\hline 
\ch{(H2O)3050} & 18300 & 36600 & 24400 & 9150\tabularnewline
\hline 
\end{tabular}
\end{table*}

\pagebreak{}

\section{Strong scalability \ch{Si705H300}}

\begin{figure}
\includegraphics{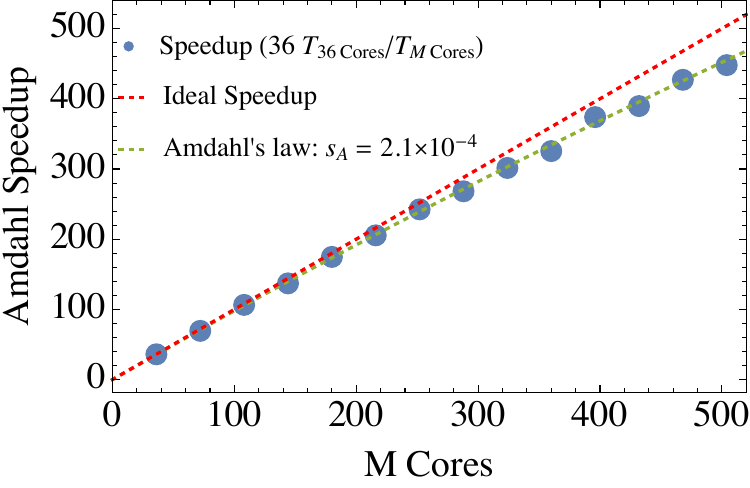}\caption{\label{fig:scalability_Si705}Strong scalability speedup analysis
for \ch{Si705H300}. The reference time for 1 processor for the speedup
is extrapolated from $T_{1}=36T_{36}$. The calculations used the
6-31G basis set (6240 basis functions) within the LDA and were performed
on several 2.60GHz Intel Xeon Gold 6240 with 256 GB using 10Gb Ethernet
networking communications.}
\end{figure}

\pagebreak{}

\section{DOS (H$_{2}$O)$_{100}$}

\begin{figure}
\includegraphics[scale=1.5]{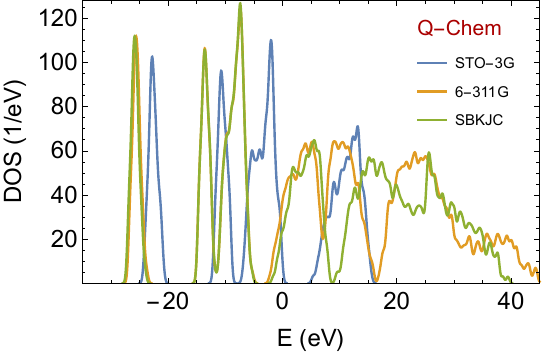}\caption{\label{fig:PPDOS}The DOS as a function of energy
for a water cluster (H$_{2}$O)$_{100}$ using the Q-CHEM\citep{shao2015advances}
program for an all-electron LDA calculation with the STO-3G and 6-311G
basis set (same data as in Figure 2.2 in the main paper, but not shifted)
and for the standard SBKJC\citep{stevens1984compact} effective core
potential (ECP) in Q-Chem, which uses a close to minimal valence basis
set. The DOS are plotted using $k_{B}T=0.01E_{h}$.}

\end{figure}

In Figure~2.2 of the main paper we show that for the
STO-3G basis set our code gives a better discription of the density
of states (DOS) in the band gap region compared to the all-electron
calculation of Q-Chem. We attribute this to the fact that our code
uses norm-conserving pseudo-potentials. In order to check this point
we ran Q-Chem with the effective core potential (ECP) SBKJC \citep{stevens1984compact},
which uses a close-to-minimal basis set and the results are given
in figure \ref{fig:PPDOS}. It can be seen that similar to the norm-conserving
pseudo-potential calculation of our code, the band gap of the SBKJC
ECP calculation is closer to the larger 6-311G basis set result. Furthermore,
it is noteworthy that the DOS in the occupied states offered by SBKJC
is nearly identical to that of the 6-311G all-electron calculation.\pagebreak{}

\section{$\textnormal{H}_{2}$ potential energy surface}

\begin{figure}
\includegraphics{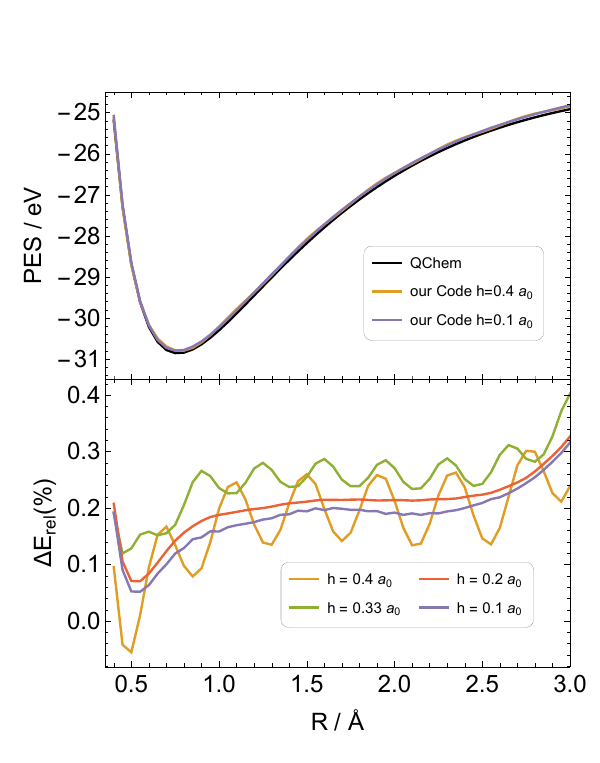}
\caption{\label{fig:PES_H2} Upper panel: The $\textnormal{H}_{2}$ molecule potential energy surface (PES) $V_{BO}\left(R\right)=E_{elec}\left(R\right)+\frac{e^{2}}{4\pi\epsilon_{0}}\frac{1}{R}$, where $R$ is the distance between the nuclei and $E_{elec}\left(R\right)$ is the DFT electronic energy. We show results calculated with our program
(for different values of the grid point spacing $h$) and with Q-Chem.
Lower panel: The relative electronic energy differences (in percent)
$\Delta E_{rel}=100\times\frac{E_{elec}^{our}-E_{elec}^{QChem}}{\left|E_{elec}^{QChem}\right|}$. }
\end{figure}

The potential energy surface (PES) is calculated in
the upper panel of Figure~\ref{fig:PES_H2} for the hydrogen molecule
with the 6-311G basis set and the local density approximation (using
the PW92 parametrization\citep{perdew1992accurate}) for our code
and for Q-Chem. For our code we chose the following additional parameters:
\begin{itemize}
\item grid length $L$: $32a_{0}$ in every cartesian direction,
chosen large enough with $L_{\zeta_{min}}\sqrt{\zeta_{min}}\gg1$
(see eq.~(\ref{WindowLength}), here $\zeta_{min}\approx0.1a_{0}^{-2}$
from the 6-311G basis set) to include the basis functions at all distances. 
\item $\eta=10^{-12}$ (see eq.~\ref{WindowLength})
\item $\beta=\infty$ (matrix diagonalization instead of
chebychev expansion for the density matrix $P$)
\end{itemize}
For the Q-Chem calculation we chose the following
specific keywords (additional to the basis set and functional selection):
\begin{itemize}
\item xc\_grid 3
\item thresh 14
\end{itemize}
Comparing our results with Q-Chem in the upper panel
of Figure~\ref{fig:PES_H2} we have overall a good agreement in absolute
energies. In the lower panel we show the PES differences between Q-Chem
and our code for different values of grid point spacing $h$. It is
noticeable that there are oscillations, also called grid corrugations,
that decrease with the grid spacing parameter $h$. This is due to
the fact, that we use the 6-311G basis set in these calculations for
which the highest Gaussian exponent is $\zeta_{max}\approx34a_{0}^{-2}$.
In general we must have $h\sqrt{\zeta_{max}}\ll1$, so that in our
case $h\ll0.17\,a_{0}$; indeed, as $h$ increases beyond this value
the corrugations increase as well. There is also a small positive
shift in all our energy calculations compared to Q-Chem's, except
for short molecular distances where there are smaller and bigger differences.
Both the shift and the sensitivity at the short bond length can be
attributed to the pseudopotentials.

Overall the approximations that we employ compared
with Q-Chem's lead to a systematic difference of $\sim0.2\text{\%}$
in the electronic energy for most of the examined distance range (and
maximally $\sim0.4\%$) and a corrugation which can be suppressed
by taking smaller grid point spacing.

\bibliography{QuadraticPaper1}